\documentclass[final]{IEEEtran} 
\IEEEoverridecommandlockouts
\usepackage{cite}
\usepackage{amsmath,amssymb,amsfonts}
\usepackage{algorithmic}
\usepackage{subcaption}
\usepackage{graphicx}
\usepackage{textcomp}
\usepackage{xcolor}
\usepackage{array}
\usepackage{hyperref}
\usepackage{etoolbox}
\appto\UrlBreaks{\do\-}
\hypersetup{draft} 

\def\BibTeX{{\rm B\kern-.05em{\sc i\kern-.025em b}\kern-.08em
    T\kern-.1667em\lower.7ex\hbox{E}\kern-.125emX}}
\begin{document}

\title{AI and ML Accelerator Survey and Trends\\
\thanks{This material is based upon work supported by the Assistant Secretary of Defense for Research and Engineering under Air Force Contract No. FA8702-15-D-0001. Any opinions, findings, conclusions or recommendations expressed in this material are those of the author(s) and do not necessarily reflect the views of the Assistant Secretary of Defense for Research and Engineering.}
}

\author{\IEEEauthorblockN{Albert Reuther, Peter Michaleas, Michael Jones, Vijay Gadepally, Siddharth Samsi, and Jeremy Kepner} \\
\IEEEauthorblockA{\textit{MIT Lincoln Laboratory Supercomputing Center} \\
Lexington, MA, USA \\
\{reuther,pmichaleas,michael.jones,vijayg,sid,kepner\}@ll.mit.edu}
}

\maketitle

\begin{abstract}

This paper updates the survey of AI accelerators and processors from past three years. This paper collects and summarizes the current commercial accelerators that have been publicly announced with peak performance and power consumption numbers. The performance and power values are plotted on a scatter graph, and a number of dimensions and observations from the trends on this plot are again discussed and analyzed. Two new trends plots based on accelerator release dates are included in this year's paper, along with the additional trends of some neuromorphic, photonic, and memristor-based inference accelerators. 

\end{abstract}


\section{Introduction}

Just as last year, the pace of new announcements, releases, and deployments of artificial intelligence (AI) and machine learning (ML) accelerators from startups and established technology companies has been modest. This is not unreasonable; for many companies that have released an accelerator report having spent three or four years researching, analyzing, designing, verifying, and validating their accelerator design trade-offs and building the software stack to program the accelerator. For those who have released subsequent versions of their accelerator, they have reported shorter development cycles, though it is still at least two or three years. The focus of these accelerators continues to be on accelerating deep neural network (DNN) models, and the application space spans from very low power embedded voice recognition and image classification to data center scale training, while the competition for defining markets and application areas continues as part of a much larger industrial and technology shift in modern computing to machine learning solutions. 

\begin{figure}[th]
    \centering
    \includegraphics[width=3in]{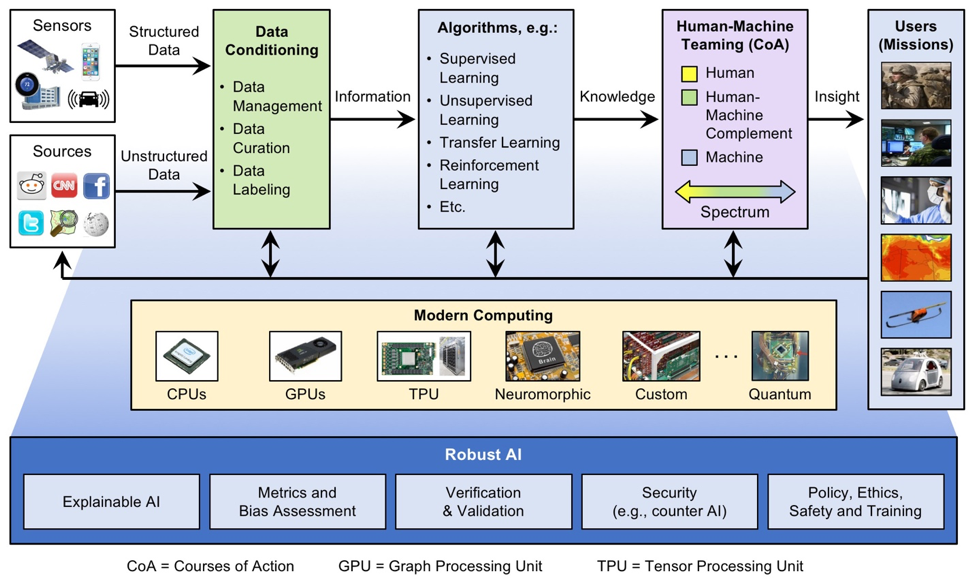}
    \caption{Canonical AI architecture consists of sensors, data conditioning, algorithms, modern computing, robust AI, human-machine teaming, and users (missions). Each step is critical in developing end-to-end AI applications and systems.}
    \label{fig:architecture}
  \end{figure}

AI ecosystems bring together components from embedded computing (edge computing), traditional high performance computing (HPC), and high performance data analysis (HPDA) that must work together to effectively provide capabilities for use by decision makers, warfighters, and analysts~\cite{gadepally2019enabling}. 
Figure~\ref{fig:architecture} captures an architectural overview of such end-to-end AI solutions and their components. 
On the left side of Figure~\ref{fig:architecture}, structured and unstructured data sources provide different views of entities and/or phenomenology. 
These raw data products are fed into a data conditioning step in which they are fused, aggregated, structured, accumulated, and converted into information. The information generated by the data conditioning step feeds into a host of supervised and unsupervised algorithms such as neural networks, which extract patterns, predict new events, fill in missing data, or look for similarities across datasets, thereby converting the input information to actionable knowledge. This actionable knowledge is then passed to human beings for decision-making processes in the human-machine teaming phase. The phase of human-machine teaming provides the users with useful and relevant insight turning knowledge into actionable intelligence or insight. 

Underpinning this system are modern computing systems. Moore's law trends have ended~\cite{theis2017end}, as have a number of related laws and trends including Denard's scaling (power density), clock frequency, core counts, instructions per clock cycle, and instructions per Joule (Koomey's law)~\cite{horowitz2014computing}. Taking a page from the system-on-chip (SoC) trends first seen in automotive applications, robotics, and smartphones, advancements and innovations are still progressing by developing and integrating accelerators for often-used operational kernels, methods, or functions.  These accelerators are designed with a different balance between performance and functional flexibility. This includes an explosion of innovation in deep machine learning processors and accelerators~\cite{leiserson2020theres,thompson2021decline,hennessy2019new,dally2020domain,lecun2019deep}. In this series of survey papers, we explore the relative benefits of these technologies since they are of particular importance to applying AI to domains under significant constraints such as size, weight, and power, both in embedded applications and in data centers.



\begin{figure*}[!htb]
    \centering
    \includegraphics[width=\textwidth]{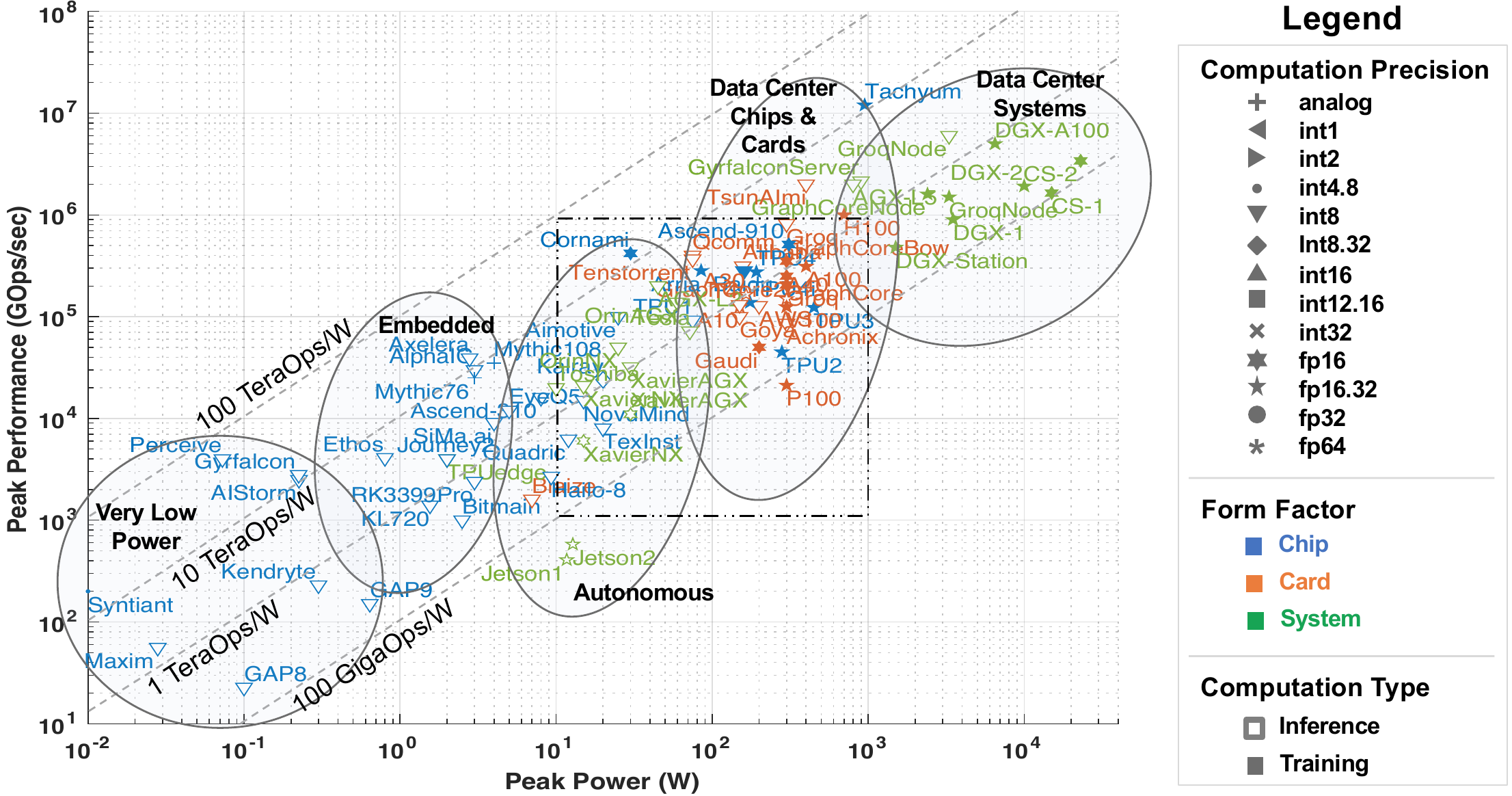}
    \caption{Peak performance vs. power scatter plot of publicly announced AI accelerators and processors.}
    \label{fig:PeakPerformancePower}
  \end{figure*}
\begin{figure}[!htb]
    \centering
    \includegraphics[width=\columnwidth]{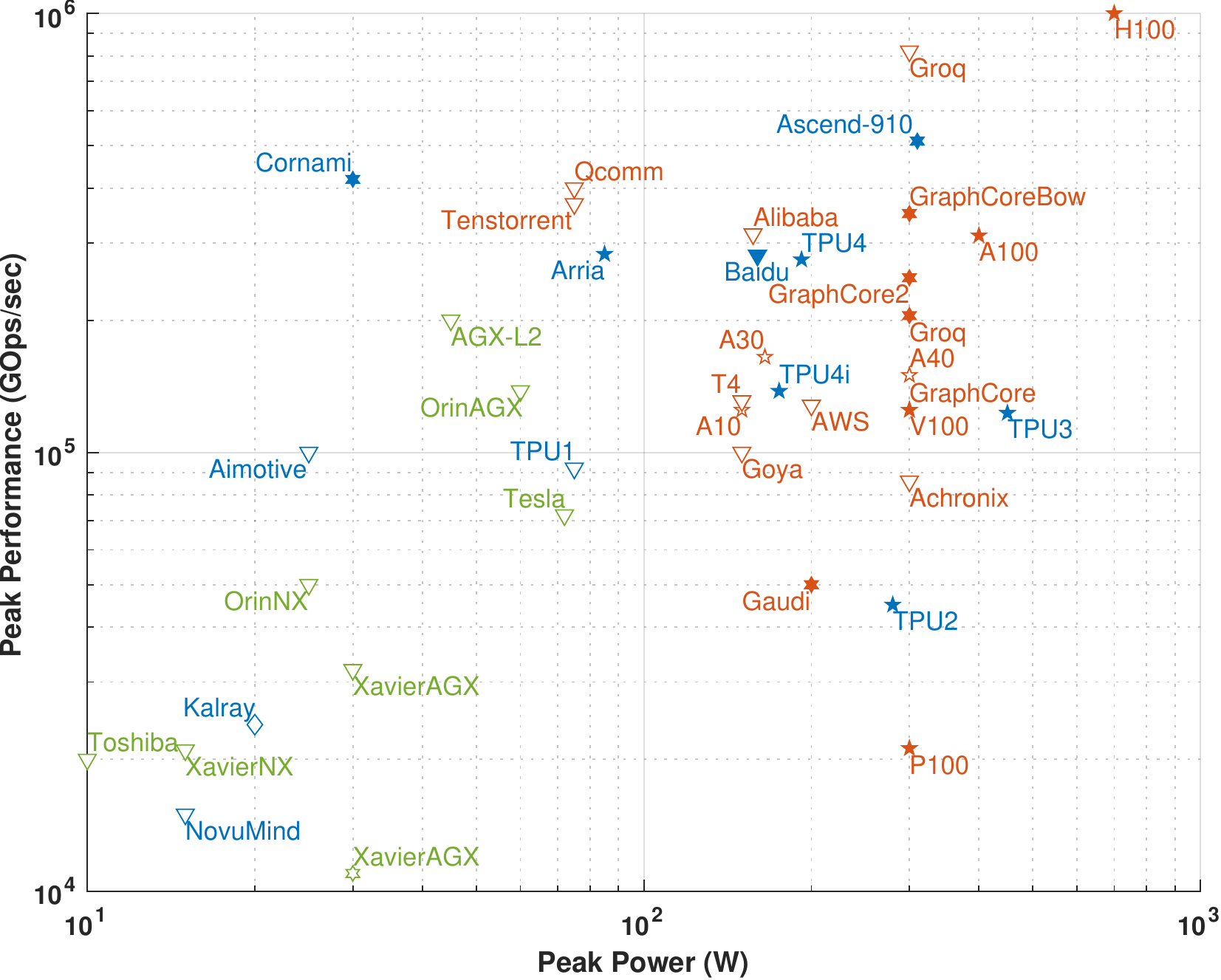}
    \caption{Zoomed region of peak performance vs. power scatter plot.}
    \label{fig:PeakPerformancePowerZoomed}
  \end{figure}

This paper is an update to IEEE-HPEC papers from the past three years~\cite{reuther2021ai,reuther2020survey,reuther2019survey}. 
As in past years, this paper continues with last year's focus on accelerators and processors that are geared toward deep neural networks (DNNs) and convolutional neural networks (CNNs) as they are quite computationally intense~\cite{canziani2016analysis}.  
This survey focuses on accelerators and processors for inference for a variety of reasons including that defense and national security AI/ML edge applications rely heavily on inference. 
And we will consider all of the numerical precision types that an accelerator supports, but for most of them, their best inference performance is in int8 or fp16/bf16 (IEEE 16-bit floating point or Google's 16-bit brain float). 

There are many surveys~\cite{lindsey1995survey,liao2001neural,misra2010artificial,sze2017efficient,sze2020efficient,langroudi2019digital,chen2020survey,wang2019deep,khan2020ai,rueckert2020digital,rogers2021academics,sunny2021survey} and other papers that cover various aspects of AI accelerators. For instance, the first paper in this multi-year survey included the peak performance of FPGAs for certain AI models; however, several of the aforementioned surveys cover FPGAs in depth so they are no longer included in this survey. This multi-year survey effort and this paper focus on gathering a comprehensive list of AI accelerators with their computational capability, power efficiency, and ultimately the computational effectiveness of utilizing accelerators in embedded and data center applications. Along with this focus, this paper mainly compares neural network accelerators that are useful for government and industrial sensor and data processing applications. A few accelerators and processors that were included in previous years' papers have been left out of this year's survey. They have been dropped because they have been surpassed by new accelerators from the same company, they are no longer offered, or they are no longer relevant to the topic. 


\section{Survey of Processors}

Many recent advances in AI can be at least partly credited to advances in computing hardware~\cite{krizhevsky2012imagenet,jouppi2018domain,hennessy2019new,dally2020domain}, enabling computationally heavy machine-learning algorithms and in particular DNNs. This survey gathers performance and power information from publicly available materials including research papers, technical trade press, company benchmarks, etc. While there are ways to access information from companies and startups (including those in their silent period), this information is intentionally left out of this survey; such data will be included in this survey when it becomes publicly available. The key metrics of this public data are plotted in Figure~\ref{fig:PeakPerformancePower}, which graphs recent processor capabilities (as of July 2022) mapping peak performance vs. power consumption. The dash-dotted box depicts the very dense region that is zoomed in and plotted in Figure~\ref{fig:PeakPerformancePowerZoomed}. 

The x-axis indicates peak power, and the y-axis indicate peak giga-operations per second (GOps/s), both on a logarithmic scale. The computational precision of the processing capability is depicted by the geometric shape used; the computational precision spans from analog and single-bit int1 to four-byte int32 and two-byte fp16 to eight-byte fp64. The precisions that show two types denotes the precision of the multiplication operations on the left and the precision of the accumulate/addition operations on the right (for example, fp16.32 corresponds to fp16 for multiplication and fp32 for accumulate/add). The form factor is depicted by  color, which shows the package for which peak power is reported. Blue corresponds to a single chip; orange corresponds to a card; and green corresponds to entire systems (single node desktop and server systems). This survey is limited to single motherboard, single memory-space systems. Finally, the hollow geometric objects are peak performance for inference-only accelerators, while the solid geometric figures are performance for accelerators that are designed to perform both training and inference. 

The survey begins with the same scatter plot that we have compiled for the past three years. As we did last year, to save space, we have summarized some of the important metadata of the accelerators, cards, and systems in Table~\ref{tab:acceleratorlist}, including the label used in Figure~\ref{fig:PeakPerformancePower} for each of the points on the graph; many of the points were brought forward from last year's plot, and some details of those entries are in~\cite{reuther2021ai}. There are several additions which we will cover below. 
In Table~\ref{tab:acceleratorlist}, most of the columns and entries are self explanatory. However, there are two Technology entries that may not be: dataflow and PIM. Dataflow processors are custom-designed processors for neural network inference and training. Since neural network training and inference computations can be entirely deterministically laid out, they are amenable to dataflow processing in which computations, memory accesses, and inter-ALU communications actions are explicitly/statically programmed or ``placed-and-routed'' onto the computational hardware. Processor in memory (PIM) accelerators integrate processing elements with memory technology. Among such PIM accelerators are those based on an analog computing technology that augments flash memory circuits with in-place analog multiply-add capabilities. Please refer to the references for the Mythic and Gyrfalcon accelerators for more details on this innovative technology. 

Finally, a reasonable categorization of accelerators follows their intended application, and the five categories are shown as ellipses on the graph, which roughly correspond to performance and power consumption: Very Low Power for speech processing, very small sensors, etc.; Embedded for cameras, small UAVs and robots, etc.; Autonomous for driver assist services, autonomous driving, and autonomous robots; Data Center Chips and Cards; and Data Center Systems.

\begingroup
\setlength{\tabcolsep}{2pt} 

\begin{table}
  \centering
  \tiny
  \caption{List of accelerator labels for plots.}
  \label{tab:acceleratorlist}

\begin{tabular}{| l | l | c | c | c | c |} \hline 

 \textbf{Company} & \textbf{Product} & \textbf{Label} & \textbf{Technology} & \textbf{Form Factor} & \textbf{References} \\ \hline 
     Achronix & VectorPath S7t-VG6 & Achronix & dataflow & Card & \cite{roos2019fpga}  \\ \hline  
     Aimotive & aiWare3 & Aimotive & dataflow & Chip & \cite{aimotive2018aiware3}  \\ \hline  
     AIStorm	 & AIStorm & AIStorm & dataflow & Chip & \cite{merritt2019aistorm}  \\ \hline  
     Alibaba & Alibaba & Alibaba & dataflow & Card & \cite{peng2019alibaba}  \\ \hline  
     AlphaIC & RAP-E & AlphaIC & dataflow & Chip & \cite{clarke2018indo}  \\ \hline  
     Amazon & Inferentia & AWS & dataflow & Card & \cite{hamilton2018aws,cloud2020deep}  \\ \hline  
     ARM & Ethos N77 & Ethos & dataflow & Chip & \cite{schor2020arm}  \\ \hline  
     Axelera & Axelera Test Core & Axelera & dataflow & Chip & \cite{ward2022axelera}  \\ \hline  
     Baidu & Baidu Kunlun 818-300 & Baidu & dataflow & Chip & \cite{ouyang2021kunlun,merritt2018baidu,duckett2018baidu}  \\ \hline  
     Bitmain & BM1880 & Bitmain & dataflow & Chip & \cite{wheeler2019bitmain}  \\ \hline  
     Blaize & El Cano & Blaize & dataflow & Card & \cite{demler2020blaize}  \\ \hline  
     Canaan & Kendrite K210 & Kendryte & CPU & Chip & \cite{gwennap2019kendryte}  \\ \hline  
     Cerebras & CS-1 & CS-1 & dataflow & System & \cite{hock2019introducing}  \\ \hline  
     Cerebras & CS-2 & CS-2 & dataflow & System & \cite{trader2021cerebras}  \\ \hline  
     Cornami & Cornami & Cornami & dataflow & Chip & \cite{cornami2020cornami}  \\ \hline  
     Enflame & Cloudblazer T10 & Enflame & CPU & Card & \cite{clarke2019globalfoundries}  \\ \hline  
     Google & TPU Edge & TPUedge & dataflow & System & \cite{tpu2019edge}  \\ \hline  
     Google & TPU1 & TPU1 & dataflow & Chip & \cite{jouppi2020domain,teich2018tearing}  \\ \hline  
     Google & TPU2 & TPU2 & dataflow & Chip & \cite{jouppi2020domain,teich2018tearing}  \\ \hline  
     Google & TPU3 & TPU3 & dataflow & Chip & \cite{jouppi2021ten,jouppi2020domain,teich2018tearing}  \\ \hline  
     Google & TPU4i & TPU4i & dataflow & Chip & \cite{jouppi2021ten}  \\ \hline  
     Google & TPU4 & TPU4 & dataflow & Chip & \cite{peckham2022google}  \\ \hline  
     GraphCore & C2 & GraphCore & dataflow & Card & \cite{gwennap2020groq,lacey2017preliminary}  \\ \hline  
     GraphCore & C2 & GraphCoreNode & dataflow & System & \cite{graphcore2020dell}  \\ \hline  
     GraphCore & Colossus Mk2 & GraphCore2 & dataflow & Card & \cite{ward2020graphcore}  \\ \hline  
     GraphCore & Bow-2000 & GraphCoreBow & dataflow & Card & \cite{tyson2022graphcore}  \\ \hline  
     GreenWaves & GAP8 & GAP8 & dataflow & Chip & \cite{greenwaves2020gap,turley2020gap9}  \\ \hline  
     GreenWaves & GAP9 & GAP9 & dataflow & Chip & \cite{greenwaves2020gap,turley2020gap9}  \\ \hline  
     Groq & Groq Node & GroqNode & dataflow & System & \cite{hemsoth2020groq}  \\ \hline  
     Groq & Tensor Streaming Processor & Groq & dataflow & Card & \cite{gwennap2020groq,abts2020think}  \\ \hline  
     Gyrfalcon & Gyrfalcon & Gyrfalcon & PIM & Chip & \cite{ward2019gyrfalcon}  \\ \hline  
     Gyrfalcon & Gyrfalcon & GyrfalconServer & PIM & System & \cite{hpcwire2020solidrun}  \\ \hline  
     Habana & Gaudi & Gaudi & dataflow & Card & \cite{gwennap2019habanagaudi,medina2020habana}  \\ \hline  
     Habana & Goya HL-1000 & Goya & dataflow & Card & \cite{gwennap2019habanagoya,medina2020habana}  \\ \hline  
     Hailo & Hailo & Hailo-8 & dataflow & Chip & \cite{ward2019details}  \\ \hline  
     Horizon Robotics & Journey2 & Journey2 & dataflow & Chip & \cite{horizon2020journey}  \\ \hline  
     Huawei HiSilicon & Ascend 310 & Ascend-310 & dataflow & Chip & \cite{huawei2020ascend310}  \\ \hline  
     Huawei HiSilicon & Ascend 910 & Ascend-910 & dataflow & Chip & \cite{huawei2020ascend910}  \\ \hline  
     Intel & Arria 10 1150 & Arria & FPGA & Chip & \cite{abdelfattah2018dla,hemsoth2018intel}  \\ \hline  
     Intel & Mobileye EyeQ5 & EyeQ5 & dataflow & Chip & \cite{demler2020blaize}  \\ \hline  
     Kalray & Coolidge & Kalray & manycore & Chip & \cite{dupont2019kalray, clarke2020nxp}  \\ \hline  
     Kneron & KL720 & KL720 & dataflow & Chip & \cite{ward2021kneron}  \\ \hline  
     Maxim & Max 78000 & Maxim & dataflow & Chip & \cite{ward2020maxim,jani2021maxim,clay2022benchmarking}  \\ \hline  
     Mythic & M1076 & Mythic76 & PIM & Chip & \cite{ward2021mythic,hemsoth2018mythic,fick2018mythic}  \\ \hline  
     Mythic & M1108 & Mythic108 & PIM & Chip & \cite{ward2021mythic,hemsoth2018mythic,fick2018mythic}  \\ \hline  
     NovuMind & NovuTensor & NovuMind & dataflow & Chip & \cite{freund2019novumind,yoshida2018novumind}  \\ \hline  
     NVIDIA & Ampere A10 & A10 & GPU & Card & \cite{morgan2021nvidia}  \\ \hline  
     NVIDIA & Ampere A100 & A100 & GPU & Card & \cite{krashinsky2020nvidia}  \\ \hline  
     NVIDIA & Ampere A30 & A30 & GPU & Card & \cite{morgan2021nvidia}  \\ \hline  
     NVIDIA & Ampere A40 & A40 & GPU & Card & \cite{morgan2021nvidia}  \\ \hline  
     NVIDIA & DGX Station & DGX-Station & GPU & System & \cite{alcorn2017nvidia}  \\ \hline  
     NVIDIA & DGX-1 & DGX-1 & GPU & System & \cite{alcorn2017nvidia,cutress2018nvidias}  \\ \hline  
     NVIDIA & DGX-2 & DGX-2 & GPU & System & \cite{cutress2018nvidias}  \\ \hline  
     NVIDIA & DGX-A100 & DGX-A100 & GPU & System & \cite{campa2020defining}  \\ \hline  
     NVIDIA & H100 & H100 & GPU & Card & \cite{smith2022nvidia}  \\ \hline  
     NVIDIA & Jetson AGX Xavier & XavierAGX & GPU & System & \cite{smith2019nvidia}  \\ \hline  
     NVIDIA & Jetson NX Orin & OrinNX & GPU & System & \cite{funk2022nvidia,nvidia2022embedded}  \\ \hline  
     NVIDIA & Jetson AGX Orin & OrinAGX & GPU & System & \cite{funk2022nvidia,nvidia2022embedded}  \\ \hline  
     NVIDIA & Jetson TX1 & Jetson1 & GPU & System & \cite{franklin2017nvidia}  \\ \hline  
     NVIDIA & Jetson TX2 & Jetson2 & GPU & System & \cite{franklin2017nvidia}  \\ \hline  
     NVIDIA & Jetson Xavier NX & XavierNX & GPU & System & \cite{smith2019nvidia}  \\ \hline  
     NVIDIA & DRIVE AGX L2 & AGX-L2 & GPU & System & \cite{hill2020nvidia}  \\ \hline  
     NVIDIA & DRIVE AGX L5 & AGX-L5 & GPU & System & \cite{hill2020nvidia}  \\ \hline  
     NVIDIA & Pascal P100 & P100 & GPU & Card & \cite{pascal2018nvidia,smith201816gb}  \\ \hline  
     NVIDIA & T4 & T4 & GPU & Card & \cite{kilgariff2018nvidia}  \\ \hline  
     NVIDIA & Volta V100 & V100 & GPU & Card & \cite{volta2019nvidia,smith201816gb}  \\ \hline  
     Perceive & Ergo & Perceive & dataflow & Chip & \cite{mcgregor2020perceive}  \\ \hline  
     Preferred Networks & MN-3 & Preferred-MN-3 & multicore & Card & \cite{preferred2020mncore, cutress2019preferred}  \\ \hline  
     Quadric & q1-64 & Quadric & dataflow & Chip & \cite{firu2019quadric}  \\ \hline  
     Qualcomm & Cloud AI 100 & Qcomm & dataflow & Card & \cite{ward2020qualcomm,mcgrath2019qualcomm}  \\ \hline  
     Rockchip & RK3399Pro & RK3399Pro & dataflow & Chip & \cite{rockchip2018rockchip}  \\ \hline  
     SiMa.ai & SiMa.ai & SiMa.ai & dataflow & Chip & \cite{gwennap2020machine}  \\ \hline  
     Syntiant & NDP101 & Syntiant & PIM & Chip & \cite{mcgrath2018tech,merritt2018syntiant}  \\ \hline  
     Tachyum & Prodigy & Tachyum & CPU & Chip & \cite{shilov2022tachyum}  \\ \hline  
     Tenstorrent & Tenstorrent & Tenstorrent & multicore & Card & \cite{gwennap2020tenstorrent}  \\ \hline  
     Tesla & Tesla Full Self-Driving Computer & Tesla & dataflow & System & \cite{talpes2020compute,wikichip2020fsd}  \\ \hline  
     Texas Instruments & TDA4VM & TexInst & dataflow & Chip & \cite{ward2020ti,ti2021tda4vm,demler2020ti}  \\ \hline  
     Toshiba & 2015 & Toshiba & multicore & System & \cite{merritt2019samsung}  \\ \hline  
     Untether & TsunAImi & TsunAImi & PIM & Card & \cite{gwennap2020untether}  \\ \hline  

\end{tabular}
\end{table}
\endgroup

For most of the accelerators, their descriptions and commentaries have not changed since last year so please refer to last two years' papers for descriptions and commentaries. There are, however, several new releases that were not covered by past papers that are covered here. 

\begin{itemize}
\item Acelera, a Dutch embedded system startup, reported the results of an embedded test chip that they have produced~\cite{ward2022axelera}. They claim both digital and analog design capabilities, and this test chip was made to test the extent of the digital design capabilities. They expect to add analog (probably flash) design elements in upcoming efforts. 

\item Maxim Integrated has released a system-on-chip (SoC) for ultra low power applications called the MAX78000~\cite{ward2020maxim,jani2021maxim,clay2022benchmarking}, which includes an ARM CPU core, a RISC-V CPU core and an AI accelerator. The ARM core is for quick prototyping and code reuse, while the RISC-V core is included to enable optimizing for the lowest power utilization. The AI accelerator has 64 parallel processors and support 1-bit, 2-bit, 4-bit, and 8-bit integer operations. The SoC operates at a maximum of 30mW and is intended for low-latency, battery-powered applications. 

\item Tachyum came out of startup stealth mode in 2017, and they just recently announced the release of an evaluation board for their Prodigy all-in-one processor~\cite{hilson2022startup}. They are promising the functionality of CPUs and GPUs within each core, and it is designed for HPC and machine learning applications. The chip is reported to have ``128 high-performance unified cores'' running at 5.7 GHz~\cite{shilov2022tachyum}. 

\item NVIDIA announced their next generation GPU called Hopper (H100) in March 2022~\cite{smith2022nvidia}. It features even more Symmetric Multiprocessors (SIMD and Tensor cores), 50\% higher memory bandwidth, and a 700W power budget for the SXM mezzanine card instance. (PCIe card power budget is 450W. 

\item Over the past couple of years, NVIDIA has also announced and released several system platforms for automotive, robotic, and other embedded applications that deploy Ampere-generation GPU architecture. Specifically for automotive applications, the DRIVE AGX platform added two new systems: the DRIVE AGX L2 that enables Level 2 autonomous driving within a 45W power envelope and the DRIVE AGX L5 that is intended to enable Level 5 autonomous driving within an 800W power envelope~\cite{hill2020nvidia}. Similarly, the Jetson AGX Orin and Jetson NX Orin also use an Ampere-generation GPU, and are intended for robotics, factory automation, etc.~\cite{funk2022nvidia,nvidia2022embedded}, and they consume a maximum of 60W and 25W peak power. 

\item Graphcore shared rough peak performance numbers for their second generation accelerator chip, the CG200~\cite{ward2020graphcore,toon2020introducing,lunden2020graphcore}. Since it is deployed on a PCIe card, we can assume that the peak power is around 300W. In the past year, Graphcore also announced it's Bow accelerator, which is the first wafer-on-wafer processor designed in cooperation with TSMC. The accelerator itself is the same CG200 as mentioned above, but it is mated with a second wafer that greatly improves power and clock distribution throughout the CG200 chip~\cite{tyson2022graphcore}. This translates into 40\% better performance and 16\% better performance-per-Watt. 

\item Almost a year after Google announced details of their fourth generation inference-only  TPU4i accelerator in June 2021~\cite{jouppi2021ten}, Google shared details about their fourth generation training accelerator, TPUv4. Very few details were announced, but they did share peak power and performance numbers~\cite{peckham2022google}. As with previous TPU variants, TPU4 is available through the Google Compute Cloud and for internal operations. 

\end{itemize}

Next, we must mention  accelerators that do not appear on Figure~\ref{fig:PeakPerformancePower} yet. Each has been released with some benchmark results but either no peak performance numbers or no peak power numbers. 

\begin{itemize}
\item After last year releasing some impressive benchmark results for their reconfigurable AI accelerator technology~\cite{ward2020sambanova} and this year publishing two deeper technology reveals~\cite{prabhakar2022sambanova,prabhakar2021sambanova} and an applications paper with Argonne National Laboratory~\cite{emani2021accelerating}, SambaNova still has not provided any details from which we can estimate peak performance or power consumption of their solutions. 

\item In May 2022, Intel's Habana Labs announced the second generations of the Goya inference accelerator and Gaudi training accelerator, named Greco and Gaudi2, respectively~\cite{peckham2022intel, morgan2022intel}. Both promised multiple times better performance than their predecessor. Greco will be a single-width PCIe card drawing 75W, while the Gaudi2 will continue to be a double-width PCIe card drawing 650W (likely on a PCIe 5.0 slot). Habana released some benchmarking comparisons to Nvidia A100 GPUs for the Gaudi2, but peak performance numbers were not disclosed for either of these accelerators. 

\item Esperanto has produced a few demo chips for evaluation by Samsung and other partners~\cite{martin2022samsung}. The chip is reported to be a 1,000-core RISC-V processor with each core having an AI tensor accelerator. Esperanto has published a few relative performance metrics~\cite{freund2021esperanto}, but they have not disclosed any peak power or peak performance values. 

\item During the Tesla AI Day event, Telsa gave some details of their custom-built Dojo accelerator and system. They did provide peak performance of 22.6 TF FP32 performance per chip, but they did not report peak power draw per chip. Perhaps these details will come later~\cite{peckham2021enter}.  

\end{itemize}

Finally there is one departure to the report this year. Last year, Centaur Technology announced a x86 CPU with an integrated AI accelerator, which was realized as a 4,096 byte-wide SIMD unit. The performance estimates were competitive, but VIA Technologies, the parent company of Centaur, sold off the USA-based engineering team of the processor to Intel, Corp. and seems to have ended the development of the CNS processor~\cite{shilov2021via}. 


\section{Observations and Trends}

There are several observations comments for us to appreciate on Figure~\ref{fig:PeakPerformancePower}. 
\begin{itemize}
\item Int8 continues to be the default numerical precision for embedded, autonomous and data center inference applications. This precision is adequate for most AI/ML applications with a reasonable number of classes. However, some accelerators also use fp16 and/or bf16 for inference. For training, has become integer representations 
\item Among the very low power chips, what is not captured is the other features beyond the machine learning accelerator on the chip. It is very common in this category and the Embedded category to release system-on-chip (SoC) solutions, which often include low-power CPU cores, audio and video analog-to-digital converters (ADCs), encryption engines, network interfaces, etc. These additional features of the SoCs do not change the peak performance metric, but they do have a direct impact on the peak power reported for the chip, so please keep this in mind when comparing them. 
\item Not much has changed in the Embedded segment, which probably means that the computational performance and peak power is adequate for the types of applications in this area. 
\item The density has become very crowded in the Autonomous and Data Center Chips and Cards segments, which required the zoomed in Figure~\ref{fig:PeakPerformancePowerZoomed}. Over the past few years, several established embedded computing microelectronics companies including Texas Instruments have released AI accelerators, while NVIDIA has released and announced several more powerful automotive and robotics application systems as mentioned above. Among the Data Center Cards, the PCIe v5 specification is highly anticipated so as to break through the 300W power limit of PCIe v4. 
\item Finally, the high-end training systems are not only posting very impressive performance numbers, but those companies have also been announcing highly scalable inter-networking technologies to network thousands of cards together. This is particularly important for dataflow accelerators like Cerebras, GraphCore, Groq, Tesla Dojo, and SambaNova, which are explicitly/statically programmed or ``placed-and-routed'' onto the computational hardware. It enables these accelerators to accommodate extremely large models like transformers~\cite{vaswani2017attention}. 
\end{itemize}

\subsection{Broader Trends}

\begin{figure}
\centering
\begin{subfigure}{0.5\textwidth}
    \includegraphics[width=\columnwidth]{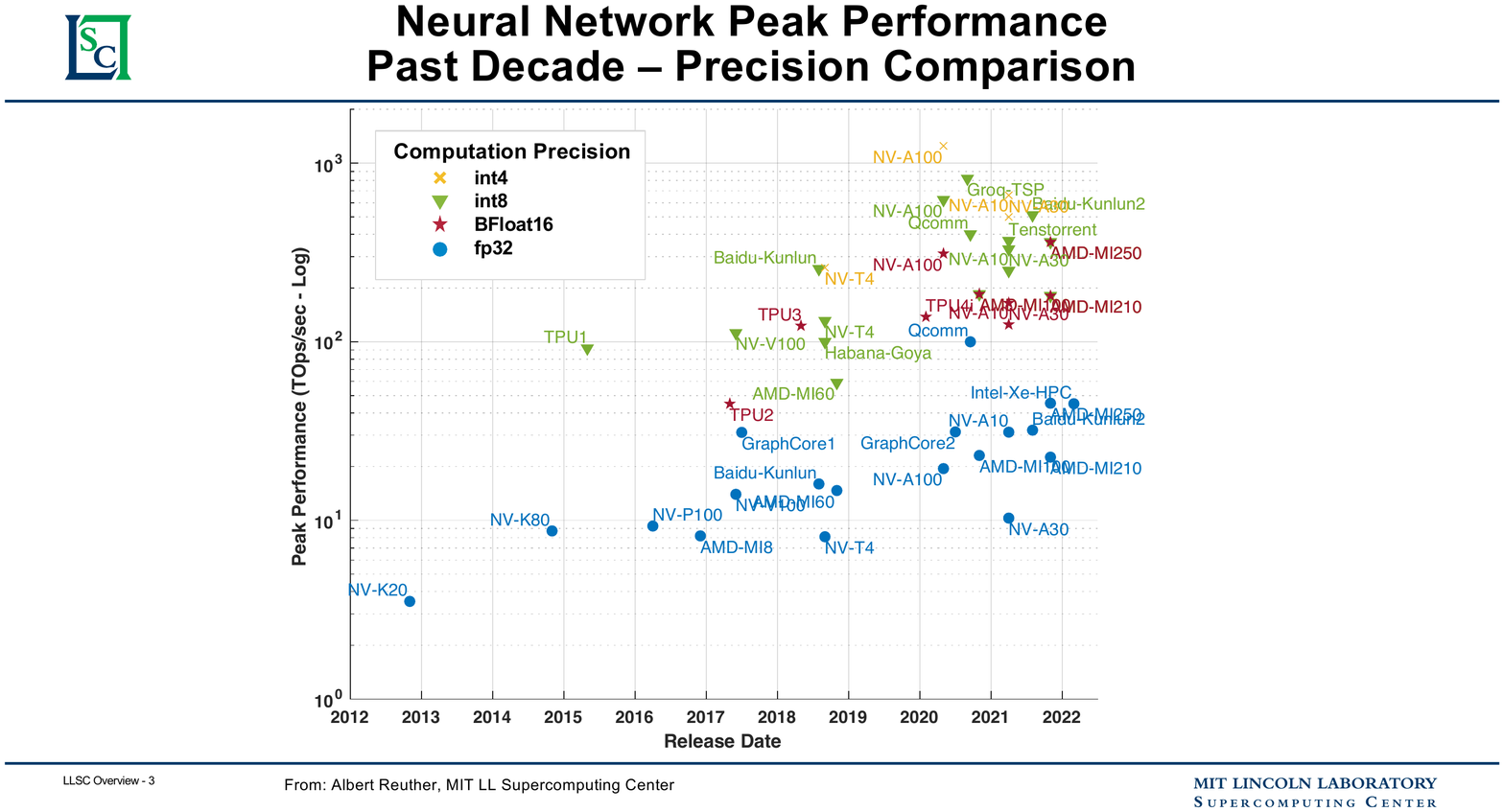}
    \caption{Peak performance for various precisions vs. release date.}
    \label{subfig:Precision}
  \end{subfigure}
\hfill
\begin{subfigure}{0.5\textwidth}
    \includegraphics[width=\columnwidth]{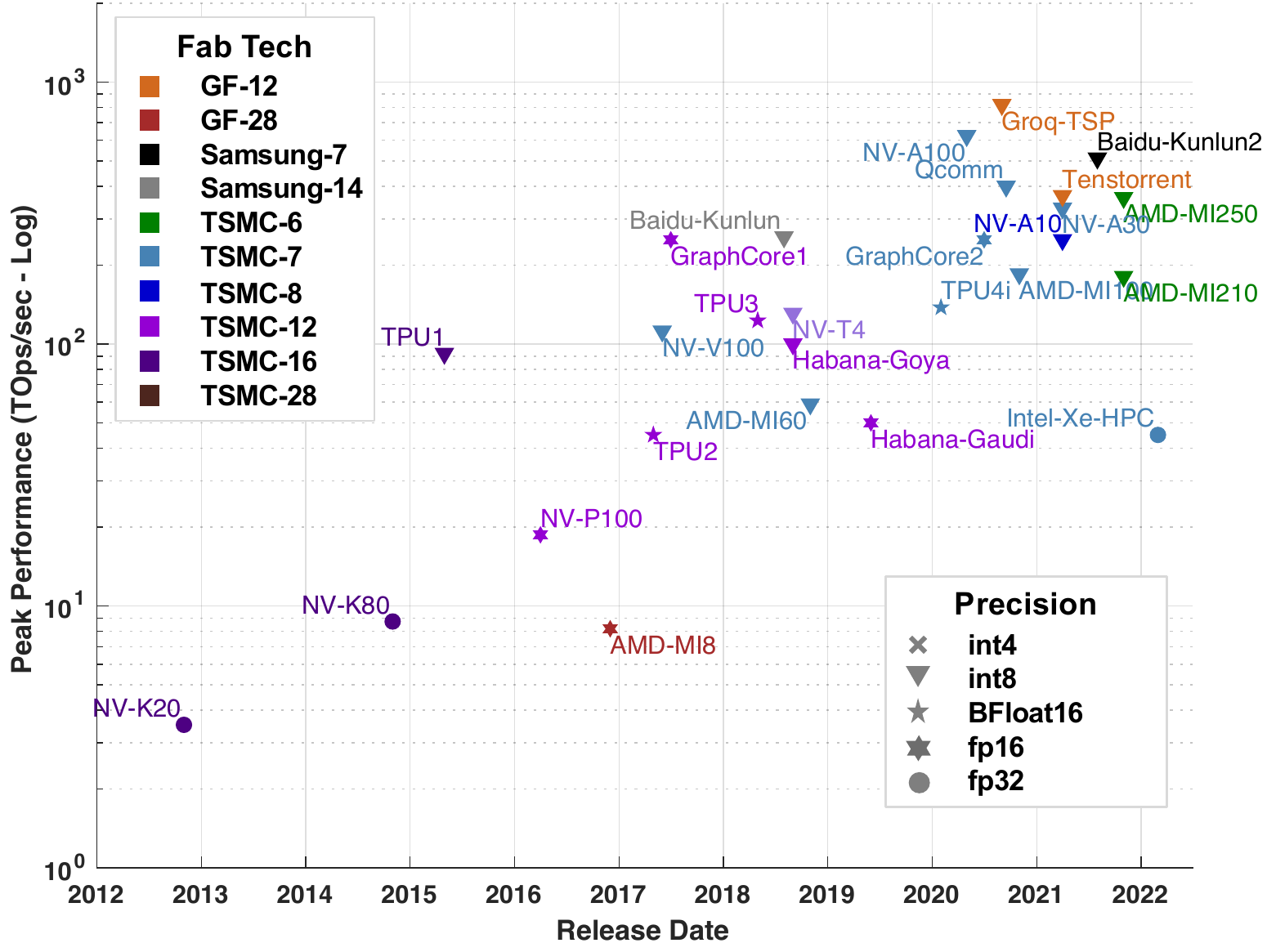}
     \caption{Peak performance and fabrication technology vs. release date.}
    \label{subfig:FabTech}
  \end{subfigure}

\caption{Trends with respect to release date for subset of publicly announced AI accelerators and processors.}
\label{fig:releasedate}
\end{figure}

We also collected release dates, fabrication technology, and peak performance for multiple precisions for a smaller subset of accelerators listed in Table~\ref{tab:acceleratorlist}. We were curious about the trends of peak performance over the past ten years and how numerical precision and fabrication technology influenced it. These data are plotted in Figure~\ref{fig:releasedate}. Figure~\ref{subfig:Precision} plots the release date of a number of accelerators versus their peak performance for one or more precision formats. There are marked gains in peak performance for each of the precision formats, but within each format the maximum gain is 1.5 orders of magnitude over the 10-year period. In Figure~\ref{subfig:FabTech}, we plot the release date versus the fabrication technology used for the accelerator. The default precision for the peak performance values is int8; however, there are a number of accelerators (e.g., NVIDIA K20, K80 and AMD Mi8) which did not have int8 support. For these accelerators, the peak performance is reported for the lowest precision that the accelerator supported. This plot shows that much performance has been gained over the past ten years by supporting lower precision formats; it is particularly interesting to observe how support for lower precision formats was included in these accelerators as research and industry explore the effectiveness of lower floating point and integer formats in CNN/DNN inference and training. 

We have several more observations and trends that are not yet captured in graphs. First, the exploration for the best numerical formats for inference and training continue. For inference, some discussion continues whether int4 will be acceptable for embedded inference, and the Maxim MAX 78000 SoC solution supports 1-bit, 2-bit, 4-bit, and 8-bit integer weights~\cite{jani2021maxim}. On the training side, it has been announced that NIVIDA Hopper, Intel Gaudi2 and a future GraphCore accelerator will support the lower precision FP8 numerical format~\cite{burt2022chip}. GraphCore posted an analysis paper on FP8~\cite{noune2022bit}, including trade-off analyses of scaled integer versus floating point representations, different 8-bit floating point representations, and mixed representation DNN model performance. 


Another trend that has caught our attention is that mathematical kernels other than DNN/CNN models have been implemented on several dataflow accelerators. These dataflow accelerators generally handle each data item independently (i.e., there are no cache lines), and data movement and computational operations are explicitly/statically programmed or ``placed-and-routed'' onto the computational hardware (as mentioned previously). Hence, they are amenable to implementing other mathematical kernels for digital signal processing, physical simulation like computational fluid dynamics and weather simulation, and massive graph processing. Cerebras demonstrated the mapping of fast stencil-code onto their wafer-scale processor~\cite{rocki2020fast}, while researchers from the University of Bristol demonstrated stencil codes and image processing using a GraphCore IPU~\cite{louw2021using}. A team from Citadel Enterprise America also reported on a series of HPC microbenchmarks that they ran on GraphCore IPUs~\cite{jia2019dissecting}. 
Google Research has been very busy demonstrating their TPUs on a variety of parallel HPC applications including flood prediction~\cite{hu2022accelerating}, large scale distributed linear algebra~\cite{lewis2021large}, molecular dynamics simulation~\cite{sharma2021molecular}, fast Fourier transforms~\cite{lu2021nonuniform,lu2020large}, MRI reconstruction~\cite{lu2020accelerating}, financial Monte Carlo simulations~\cite{belletti2019tensor}, and Monte Carlo simulation of the Ising model~\cite{yang2019high}. We see this as a foreshadowing of more interesting research and development in using this high performance accelerators.


\subsection{Other Technologies}
The word neuromorphic has become a nebulous term. In industry, it seems to have settled on any computational circuit that in some way mimics some aspects of how the synapses in brains work. When this is applied most broadly, it encompasses many if not all of the accelerators that this series of papers surveys. In academia and the broader research world, neuromorphic computing is the research, design, and development of computational hardware that models functionality and processes in brains, including chemical processes and electrical processes~\cite{schuman2017survey,james2017historical}. These brain process simulation efforts have spanned the past four decades, but there is only a modest overlap with the accelerators that are captured in these surveys. 

One clear overlap is circuitry based on spiking neural networks, which is what we will focus on here. 
Intel probably has the most extensive research program for evaluating the commercial viability of spiking neural network accelerators with their Loihi technology~\cite{davies2018loihi,orchard2021efficient} and Intel Neuromorphic Development Community~\cite{davies2021advancing}. Among the applications that have been explored with Loihi are target classification in synthetic aperture radar and optical imagery~\cite{barnell2020target}, automotive scene analysis~\cite{viale2021carsnn}, and spectrogram encoder~\cite{orchard2021efficient}. 
Further, one company, Innatera, has announced a commercial spiking neural network processor~\cite{ward2021innatera}. They have shared an example inference benchmark demonstration~\cite{levy2021innateras}, but they have not release peak performance or power numbers.
In a related vein, some memristor technology is showing its effectiveness in simulating variable neuron-synapse functionality. However, the use of memristors in AI/ML accelerators is still very much in the research phase. A company call Knowm is working towards commercialization of a memristor-based accelerator~\cite{ostrovskii2022structural}, but that is probably a few years away. They do sell a memristors and an evaluation kit on their website. 

Progress continues to be made in building and commercializing silicon photonic for AI/ML accelerators, including an extensive survey paper~\cite{sunny2021survey}. Several optical/photonic startups have announced photonic inference processors, including LightMatter~\cite{ward2020optical}, Lightelligence~\cite{ward2021optical}, LightOn~\cite{launay2020light}, and Optalysys~\cite{cottle2020optical,wilson2020multiply}, and several of these companies have suggested that they will publish performance and power measurements later this year~\cite{schneider2019neural, choi2022photonic}. The LightMatter, Lightelligence, and LightOn accelerators implement multiply-accumulate computations directly with Mach-Zehnder interferometers, while the Optalysys uses an 2-dimensional FFT technique also based on Mach-Zehnder interferometers. 

\section{Summary}

This paper updated the survey of deep neural network accelerators that span from extremely low power through embedded and autonomous applications to data center class accelerators for inference and training. We focused on inference accelerators, and discussed some new additions for the year. The rate of announcements and releases has continued to be consistent and modest.

\section{Data Availability}

The data spreadsheets and references that have been collected for this study and its papers will be posted at \url{https://github.com/areuther/ai-accelerators} after they have cleared the release review process. 

\section*{Acknowledgement}

We express our gratitude to 
Masahiro Arakawa, Bill Arcand, Bill Bergeron, David Bestor, Bob Bond, Chansup Byun, Nathan Frey, Vitaliy Gleyzer, Jeff Gottschalk, Michael Houle, Matthew Hubbell, Hayden Jananthan, Anna Klein, David Martinez, Joseph McDonald, Lauren Milechin, Sanjeev Mohindra, Paul Monticciolo, Julie Mullen, Andrew Prout, Stephan Rejto, Antonio Rosa, Matthew Weiss, Charles Yee, and Marc Zissman
for their support of this work. 


\bibliographystyle{IEEEtran} 
\bibliography{MLAcceleratorTrends}


\end{document}